\documentstyle[12pt]{article}
\evensidemargin \oddsidemargin
\catcode `\@=11
\def\numberbysection
{\@addtoreset{equation}{section}
         \renewcommand
{\theequation}{\thesection.\arabic{equation}}}
\def\be{\begin{equation}}
\def\ee{\end{equation}}
\newcommand{\ba}{\begin{eqnarray}}
\newcommand{\ea}{\end {eqnarray}}
\newcommand{\nn}{\nonumber}

\def\d{\delta}

\def\eps{\epsilon}

\def\l{\lambda}

\def\s{\sigma}

\def\ra{\rightarrow}
\def\lra{\longrightarrow}

\def\inf{\infty}
\def\half{\frac{1}{2}}

\numberbysection

\def\PA {{\cal P_{\cal A}}}
\def\P {{\cal P}}
\def\Pnk {{\cal P}_{n,k}}

\def\Rem {{\bf Remark: }}

\def\lii {(l_i)_i}
\def\hii {(g_i)_i}
\def\lan {\langle}
\def\ran {\rangle}
\def\V {{V(c,h)}}
\def\tvnk {{{\widetilde V}_{n,k}}}
\def\Vnk {{{V}_{n,k}}}
\def\Vhnk {{{\widehat {V}}_{n,k}}}
\def\pnk {{\pi_{n,k}}}
\def\snk {{{s}_{n,k}}}
\def\Lhn {{\widehat L_n}}
\def\Cn {{C_n}}
\def\Lhm {{\widehat L_m}}
\def\Cm {{C_m}}
\def\N {{I \! \! N}}
\def\Z {{Z \! \! \! Z}}
\def\C {{\rm l}\!\!\!{\rm C}}
\def\hp {\hspace{\parindent}}

\begin {document}

\begin{titlepage}
\begin {flushright}
BONN-TH-94-05
hep-th/9405041
\end {flushright}
\vfil
\begin {center}

{\Large \bf PATHSPACE DECOMPOSITIONS} \\ [3mm]
{\Large \bf FOR THE VIRASORO ALGERBA} \\ [3mm]
{\Large \bf AND ITS VERMA MODULES}

\vfil

{\Large by \\[4mm]
Ralph M. Kaufmann$^*$}\\[4mm]
{\it Physikalisches Institut der Universit\"at Bonn}\\[3mm]
{\it Nussallee 12, 53115 Bonn, Germany}
\vfil

\end {center}

\begin {abstract}
Starting from a detailed analysis of the structure
 of pathspaces of the ${\cal A}$-fusion graphs
and the corresponding irreducible Virasoro algebra
quotients $V(c,h)$ for the ($2,q$ odd) models,
we introduce  the notion of an admissible pathspace
representation.  The pathspaces ${\cal P_A}$ over the
${\cal A}$-Graphs are isomorphic to the pathspaces
over Coxeter $A$-graphs that appear in FB models.
We give explicit  construction algorithms for
admissible representations. From the
finitedimensional results of these algorithms
we derive a decomposition of $V(c,h)$ into its
positive and negative definite subspaces w.r.t.
the Shapovalov form and the corresponding
signature characters. Finally, we treat the
Virasoro operation on the lattice induced by
admissible representations adopting a particle
point of view. We use this analysis to decompose
the Virasoro algebra generators themselves. This
decomposition also takes the nonunitarity of the
($2,q$) models into account.
\end {abstract}

\vfil

$^*$ kaufmann@avzw02.physik.uni-bonn.de

\end{titlepage}

\pagebreak

\section {Introduction}

\hp Ever since the paper \cite {PP}, the close
connection between CFT and statistical mechanics
has been known. Subsequently, Belavin, Polyakov and
Zamolodchikov  \cite {BPZ}  gave an in-depth
analysis of the implications of conformal
invariance in two dimensions \cite {P1,Mig,P2}.
The above mentioned correspondence can effectively
be analyzed on the level of graphs and their
pathspaces. These made their first
appearance in statisitcal mechanics as the
configuration spaces of integrable models
defined by Andrews, Forrester and Baxter
\cite {ABF}.  Since the subsequent identification
of critical parameters by Huse \cite {Huse} as
belonging to the unitary discrete FQS-series of
minimal models \cite {FQS}, there has been a desire
to construct a Virasoro representation on these
path spaces in order to better understand the
appearence of this algebra in this context. This
would be especially intriguing, since the algebra
appears even in off critical cases in the
calculation of local height probabilities
\cite {riggs}. One interesting aspect would be a
correspondence between the Temperly Lieb Jones
algebra predominant in these models \cite {pcomm}
and the Virasoro algebra itself providing a link
between statistical systems and the corresponding
conformal field theories \cite {tlj1,tlj2}.
There has been a lot of research in developing
new models along the lines of \cite {ABF}  and in
the study of their properties
\cite {pcomm,OB,Detal,pjourn,PS,DFJMN,JMO,DJK}.
In \cite {FB} Forrester and Baxter introduced
new models whose critical behavior is nonunitary
\cite {riggs}.  In this context one can find the
(2,q odd) series of minimal models in the critical
behaviour of statistical models based on Coxeter
$A_n$ graphs which will be studied in this paper.
The class of models based on Coxeter graphs can be
extended to include all A-D-E graphs \cite {pnuc}
parallel to the classification of modular invariant
partition functions \cite {class1,class2,class3,
class4}.

There are two ways in which to proceed in order to
arrive at the correspondence mentioned above. One
can start from the statistical model side or from
the CFT side. There has been some development on
the statistical side (\cite {KS} and ref. therein).
 In \cite {KS}, the authors propose a double limit
in order to obtain Virasoro generators from the
Temperly Lieb Jones algebra. For a different
approach see \cite {IT,T,FT}.

We shall tackle the above problem
(construction of a Virasoro representation) from
the CFT side where only little is known.
On the CFT side, pathspaces appear in sum formulas
for the conformal characters \cite {a.r}. These
pathspaces, however, are not over simple Coxeter
graphs, but are more involved in the sense that
the structure of the graphs which are considered
is a bit more complicated. However, the $\cal A$
graphs which appear in the (2,q) models (q odd)
can be linked via pathspace isomorphisms to the
Coxeter $A_n$ graphs \cite {a.r} (see also section
2). In \cite {KRV}, another way of introducing
graphs to CFT was found by rewriting character
formulas.

In this paper, we will proceed from the pathspaces
over the $\cal A$ graphs. In section 2 we will
give the relavant definitions and notations which
we will need in the following. Then, in section 3,
we will start the analysis of the pathspace and
the Verma module quotient structure in order to
establish a connection between them. As a result,
we will introduce the notion of an admissible
pathspace representation. In section 4, we will
give and discuss possible constructions for these
representations. The calculations involving these
constructions will then lead us to the main
conjecture which states signature chararacter
formulas for the (2,q)-models as well as the
corresponding decomposition of the irreducible
Virasoro quotient $\V$ into positive and negative
definite subspaces $\V = \V^+ \oplus \V^-$. Finally,
 in section 5, we turn to the structure of the
Virasoro action induced by admissible
representations on the path space. To this end,
a particle interpretation is given. The action of
the Virasoro generators $L_n$ can then be
described by a shift ($\Lhn$) and a one particle
creation (resp.\ annihilation) $(\Cn)$ operator.
The main conjecture can be restated in formulas
for these operators and their adjoints. This
reformulation explains the degree of nonunitarity
of the (2,q) or, in other words, the degree of
non-selfadjointness of the Virasoro algebra in the
pathspace metric. We conclude the paper with a
summary of the results and an outlook into further
fields of study.

\section {Preliminaries and Notations}

\hp Since we are interested in representations of the
Virasoro algebra, it is useful to first fix
notations.\\
By the Virasoro algebra $Vir$ we mean the complex Lie
algebra generated by $L_n, n \in {\Z}$ and $C$ with
commutation relations
\ba
\label {comm}
{} [L_n,L_m] &=& (m-n) L_{m+n} + \d_{m+n,0}
\frac {1}{12} C (m^3-m) \\
{} [L_n,C] &=&0.
\ea
Please note the choice of signs in (\ref {comm}).
As usual, we denote by $M(c,h)$ the Verma module to
the eigenvalue c of $C$ and lowest (choice of signs)
weight h for $L_0$.  Furthermore, let $J(c,h)$ be
the unique maximal submodule and $\V$ the unique
irreducible quotient:
\be
\V= M(c,h)/J(c,h).
\ee
In this paper,
we will concentrate on certain values
of $c$, namely the ones belonging to the (2,q)
series. Here $(p,q)$ denotes the $c$ value:
\be
c(p,q) = 1-6 \frac {(p-q)^2}{pq}.
\ee
For these specific models, an explicit basis for
the quotients $\V$ was given by Feigin, Nakanishi
and Ooguri (FNO) in \cite {FNO}, the lowest weights
being:
\be
h_j = \frac {-j(q-2-j)} {2q}, \hspace {2cm} j=0,
\ldots, N
\ee
with $q = 2N+3$.
The basis is given by elements
\be
L_{n_1} \cdots L_{n_m} \vert h_j \ran
\ee
with $ n_1 \geq \ldots \geq n_m \geq 1$ which
satisfy the following two conditions:
\begin {itemize}
\item [1)]  $n_i- n_{i+N} \geq 2$   (difference
two condition)
\item [2)]  $\# \{ n_i =1 \} \leq j  $   (initial
condition).
\end {itemize}
In the paper \cite {FNO}, the authors also
introduce an order which will be quite helpful
later on. One defines:
\be
L_{m_1} \cdots L_{m_r} \succ L_{n_1} \cdots L_{n_s}
\label {ord}
\ee
if
\begin {itemize}
\item [i)] $r>s$,
\item [ii)] $r=s$ and $\sum_i m_i > \sum_i n_i$ or
\item [iii)] $r=s$, $\sum_i m_i = \sum_i  n_i$ and
$m_t >n_t, \; m_i =n_i$ for $1\leq i \leq t-1$.
\end {itemize}

For the above basis a graphical enumeration method
was given in \cite {a.r}. The graphs used for this
are the fusion graphs ${\cal A}_{N+1}$ of the
corresponding theory with $c(2N+3,2)$ (for a precise definition see
\cite {a.r}~). Here we just define the graph
${\cal A}_N$ by its incidence matrix.
For $0\leq i,j \leq N-1$:
\[ ( {\cal A}_N )_{i,j} =  \left\{ \begin{array}{ll}
			1 &\mbox{if $i+j < N$}\\
			   0 & \mbox{otherwise}
			       \end {array}
\right. . \]

The basis itself is then given by paths over the
corresponding graphs. By this we mean the following:\\
A path over a graph  $\cal G$ with vertices
numbered by a set $I$ is a map : ${\N_0} \mapsto I$
i.e. a sequence of vertices $\lii$, with the
restriction that two successive vertices are linked.\\
Furthermore, let
${\P}_{\{ {\cal G},l_0,l_{\inf} \} }$ denote
the Hilbert space with basis given by the paths on
the graph $\cal G$ which start at $l_0$ and end in
$l_{\inf}$. This means that for
$i \gg 1$, $l_i = l_{\inf}$.
The scalar product on the space is chosen to be
the one in which all path are mutually orthonormal.

We then have a bijection \cite {a.r} between
${\P}_{ \{ {\cal A}_{N+1},N-j,0 \} }$ and
$V(c(2,2N+3),h_j)$ which is given by the simple map:
\be
((l_i)_i ) \mapsto \ldots L_2^{l_2}L_1^{l_1}
= \prod_{i=\inf}^1 L_i^{l_i} \hspace {1cm}
\mbox {for $i \in {\N}$}.
\label {fusiso}
\ee
The restrictions for the paths given by the graph
are directly translated into the difference two
condition of the FNO basis. The index $N+1$  fixes
$c$ to $c(2,2N+3)$ and the first term of the
sequence $l_0$
dictates the h value of the theory and guarantees
compliance with the initial condition. In the
pathspaces we have an $L_0$ which provides the
usual grading:
\be
L_0 \lii := (h_{N-l_0} +\sum_{i \in {\N}} i l_i )
\cdot (\lii).
\label {l0}
\ee
By K-theoretic arguments one can show that the
pathspace over the ${\cal A}_{N+1}$ graph is
isomorphic to the Coxter $A_{2(N+1)} = A_{q-1}$
graph considered in \cite {riggs} and \cite {FB}.

{\bf Example: } In the case of the Lee-Yang edge
singularity (the (2,5) model), an isomorphism can
easily be given. We just relabel the $A_4$ graph:\\

\begin {figure} [h]
\hspace {2cm}
\setlength{\unitlength}{0.01in}%
\begin{picture}(461,43)(60,720)
\thicklines
\put(517,753){\circle*{8}}
\put(129,753){\circle*{8}}
\put(286,753){\circle*{8}}
\put(183,753){\circle*{8}}
\put(236,753){\circle*{8}}
\put(360,753){\circle*{8}}
\put(414,753){\circle*{8}}
\put(467,753){\circle*{8}}
\put(310,753){\line( 1, 0){ 25}}
\multiput(335,753)(-0.50000,0.25000){17}
{\makebox(0.4444,0.6667){\rm .}}
\multiput(335,753)(-0.50000,-0.25000){17}
{\makebox(0.4444,0.6667){\rm .}}
\put(129,753){\line( 1, 0){ 49}}
\put(183,753){\line( 1, 0){ 49}}
\put(236,753){\line( 1, 0){ 50}}
\put(360,753){\line( 1, 0){ 49}}
\put(414,753){\line( 1, 0){ 49}}
\put(467,753){\line( 1, 0){ 50}}
\put(285,720){\makebox(0,0)[lb]
{\raisebox{0pt}[0pt][0pt]{\rm 4}}}
\put(360,720){\makebox(0,0)[lb]
{\raisebox{0pt}[0pt][0pt]{\rm 1}}}
\put(520,720){\makebox(0,0)[lb]
{\raisebox{0pt}[0pt][0pt]{\rm 1}}}
\put( 60,750){\makebox(0,0)[lb]
{\raisebox{0pt}[0pt][0pt]{{\large $A_4$:}}}}
\put(130,720){\makebox(0,0)[lb]
{\raisebox{0pt}[0pt][0pt]{\rm 1}}}
\put(185,720){\makebox(0,0)[lb]
{\raisebox{0pt}[0pt][0pt]{\rm 2}}}
\put(235,720){\makebox(0,0)[lb]
{\raisebox{0pt}[0pt][0pt]{\rm 3}}}
\put(415,720){\makebox(0,0)[lb]
{\raisebox{0pt}[0pt][0pt]{\rm 0}}}
\put(465,720){\makebox(0,0)[lb]
{\raisebox{0pt}[0pt][0pt]{\rm 0}}}
\end{picture}

\caption{Relabeling of the $A_4$ graph}
\end {figure}
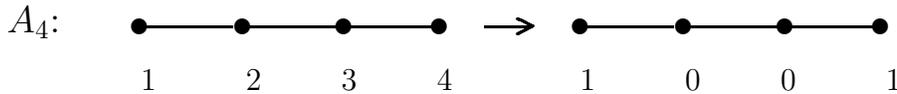

\pagebreak

and see that in this labeling the path space over
$A_4$ is just the same as over the ${\cal A}_2$
graph. \\

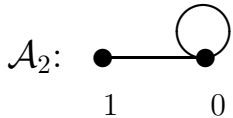
\begin {figure} [h]
\hspace {2cm}
\setlength{\unitlength}{0.0125in}%
\begin{picture}(92,45)(160,725)
\thicklines
\put(243,749){\circle*{8}}
\put(200,749){\circle*{8}}
\put(242,760){\circle{20}}
\put(200,749){\line( 1, 0){ 43}}
\put(200,725){\makebox(0,0)[lb]
{\raisebox{0pt}[0pt][0pt]{\rm 1}}}
\put(245,725){\makebox(0,0)[lb]
{\raisebox{0pt}[0pt][0pt]{\rm 0}}}
\put(160,745){\makebox(0,0)[lb]
{\raisebox{0pt}[0pt][0pt]{{\large ${\cal A}_2$:}}}}
\end{picture}

\caption {${\cal A}_2$ fusion graph}
\end {figure}

One can also check that the $L_0$ (\ref {l0})
gives the same spectrum over the ground state as
in the FB models.\\
This example shows the obvious advantages of the
$\cal A$ graphs. The ground state $(\ldots 32323
\ldots)$ of the FB model reads simply as $(000
\ldots)$ and the operator $L_0$ just depends on
the site $i$, not on three neighboring sites. We
also have that the energy of the ground state is 0
and thus it is much easier to find the energy of
the excited state, since no ground state
contribution has to be subtracted.

 In general, we can see that in the $A_{2n}$
models there are just n distinct labels which
can appear in the even resp.\ odd lattices,
thus a graph with N vertices should suffice.
Furthermore, one feature of the $\cal A$ graphs
is that the vertex 0 is always connected to itself.
So we always have a ground state $(000 \ldots)$
which replaces a ground state
$(\ldots l,l-1,l,l-1 \ldots)$ and thus we also
have a simpler structure for the excitations. In
fact, part of this has already been realized in
\cite {ABF}. The authors introduce the following
relabeling of the Coxeter $A_{r-1}$ graph, if r is
odd ($r=2N+3$):
\be
 l_i \mapsto  \left\{ \begin{array}{ll}
	  \half (2N+1- l_i) &\mbox{if $l_i$ odd} \\
	  \half (l_i-2) & \mbox{if $l_i$ even}
\end {array}
\right.
\label {lg}
\ee
These labels correspond to the ones of the fusion
graphs. Furthermore they map the initial conditions
of the fusion  and FB pathspaces for the $h$-values
onto each other. Thus one can reassociate the fields
of the given CFT to the edges of the graph in a
procedure reverse to the one presented in
\cite {a.r}.

In particular, for the $(2,5)$ model we see that
the fusion ${\cal A}_2$  and the relabeled
Coxeter $A_4$ graph are identical. In this way,
all results for the pathspaces of the former can
be translated back to the pathspace of the Coxeter
$A_4$-graph. For the translation one must, however,
first make a choice for the  odd and even
sublattices (see \cite {ABF}).

The above relabeling was interpreted in \cite {ABF}
as a lattice gas picture. For the ABF resp. FB
models, the restriction given by the graph amounts
to the restriction that the sum of particles on
neighboring lattice sites must be $N$ or $N-1$.
In this line of thinking, the fusion graphs
describe a lattice gas with the perhaps more
natural restriction that there are at most $N$
particles on two neighboring sites. The described
particle interpretation as a lattice gas is
adopted in this paper.

In the following we want to extend the operator
$L_0$ (\ref {l0}) to a full representation of $Vir$.

\section {Admissible Representations for $\PA$}
\subsection {General Situation}

\hp We are looking for an irreducible representation
of the Virasoro algebra on the pathspace
$\PA$. One other way to state this problem is
given by the following:

\Rem  If such a representation exists, then
\be
\PA \cong V(c,h) \hspace {1cm}
\mbox{as $Vir$ modules}.
\ee
Because of the universal property of the
Verma module there exists a surjective homomorphism
$\pi : M(c,h) \ra\PA$ \cite {Kac}. Then
$K:= ker(\pi)$ is a submodule. Therefore $K$ is a
submodule of the unique maximal submodule $J(c,h)$.
\\
Hence $K = J(c,h)$, since
$dim_n K := dim_n  J(c,h)$, using the isomorphism
(\ref {fusiso}).

So the problem can be reformulated in the following
way:
The object we are looking for is an isomorphism
\be
\Phi : \PA \lra V(c,h).
\label{iso}
\ee
One such isomorphism can be easily given in the
following way:

{\bf Example 1:}
\be
\Phi ((l_i)_i ) := \prod_{i=\inf}^1 L_i^{l_i}.
\label {ex}
\ee
This is possible since the $\cal A$-graphs encode
exactly the restrictions which appear for the FNO
basis.\\
Example 1, however, does not respect all of the
various natural structures of $\PA$ and $V(c,h)$
which will be introduced in the following section.

\subsection {Natural structures on $\PA$
and $V(c,h)$}

\hp If we regard $ {\PA}$ (in the following denoted by
${\P}$) simply as a configuration space of
pathspace origin, we can associate the following
structures:
First we have two operators $L_0$ and $K$ on $\P$,
where $L_0$ is the usual pathspace $L_0$
(\ref {l0}):
\be
L_0 (\lii) := (h_{N-l_0} + \sum_{i \in {\N}} i l_i )
 \cdot (\lii),
\ee
and $K$ is defined by:
\be
K (\lii) :=  (\sum_{i \in {\N}} l_i) \cdot  (\lii).
\ee
Together they provide a double grading:
\be
\P = \bigoplus_{n,k} \Pnk
\label {dgr}
\ee
where
\[
L_0 \vert_{\Pnk} = (n+h_{N-l_0}) \cdot id
\]
\[
K \vert_{\Pnk}  = k \cdot id .
\]
Furthermore, we have a natural scalar product
$( \; ,\; )_{\P}$ of the Hilbert space which takes
the configuration space nature into account. It is
defined by the fact that all paths are mutually
orthonormal:
\be
((\lii),(\hii))_\P := \prod_i  {\d_{l_i,g_i}}.
\ee
This metric turns (\ref {dgr}) into an orthogonal
decomposition
\be
\P = \perp_{n,k} \Pnk.
\ee

So the pathspace basis should yield an orthogonal
basis for the Verma module quotient under a
reasonable representation (resp. isomorphism
${\Phi} $). And we should find a structure
corresponding to the double grading (\ref {dgr}).

If, on the other hand, we look at $ V(c,h) $ we
have the usual grading given by $L_0$ and a
canonical hermitian form, the Shapovalov form
(see e.g. \cite {Kac}) which we denote by
$ \lan \; , \; \ran_S $.\\
Together, these provide the usual orthogonal
decompositon of $\V$:
\be
\V = \perp_n V_n(c,h).
\ee
Although there is no obvious second grading which
would provide another orthogonal splitting, we do
have a filtration.

Consider the spaces
\[ \tvnk  :=  span  \lan L_{n_l}
\ldots L_{n_1}  \vert n_l \geq \ldots \geq n_1, l
\leq k \mbox{ and } \sum_i n_i = n \ran \]
(i.e. linear combinations of products of at most
k $L_i $'s of energy n.)

These spaces give the above mentioned filtration
of each  $V_n(c,h)$
\be
\emptyset = \widetilde V_{n,-1} \subset
\widetilde V_{n,0} \subset \cdots \subset
\widetilde V_{n,k-1}  \subset \widetilde V_{n,k}
\subset \widetilde V_{n,k+1} \subset \cdots
\subset \widetilde V_{n,\inf}= V_n(c,h).
\label {fil}
\ee

The successive quotients of this filtration
are especially interesting, since they have
the same dimensions as the respective pathspace
counterparts. To this end
let $\Vhnk := \tvnk / \widetilde V_{n,k-1} $
denote these quotients
then we have again using (\ref {fusiso}):
\be
dim \Vhnk = dim \Pnk.
\label {dim}
\ee

We denote by ${\pnk }$ be the canonical projection
$\pnk: \tvnk \lra \Vhnk$.

After these considerations it is clear which
special features an isomorphism (\ref{iso}) should
have in order to preserve the natural structures
of $\P$.
The following definition is therefore central.

We call an isomorphism $\Phi : \PA \lra V(c,h)$
{\bf admissible}, if the following two conditions
hold:
\begin{itemize}
\item [a)] $\Phi (\Pnk) \subset \tvnk$
and thus $\pnk \circ\Phi (\Pnk) = \Vnk$.
\item [b)] $ \Phi^* \lan $   ,   $\ran_S $ is
orthogonal in the path-basis \\
i.e. $ \lan \Phi (\lii),\Phi (\hii) \ran
= \eps  (\lii,\hii) \cdot (\lii,\hii)_\P$ ,
with $\eps  (\lii,\hii) \in {\C}$.
\end {itemize}

\pagebreak

{\bf Remarks:}
\begin {itemize}
\item [1)] We can always normalize so that the
$\eps$ of b) is $\pm 1$.
\item [2)] The remark in condition a) encodes the
information (\ref {dim}) about the dimensions of
the corresponding spaces. Condition a) itself
coarsely identifies the number of $L$ 's with the
sum of the nonzero entries in the path sequence.
In particular, an entry k signifies the presence of
k $L$ 's (except for $l_0$, of course, which just
fixes h).
\item [3)] Condition b) guarantees the existence of
an orthogonal basis complementing the previous one
in each step of the filtration. \\
Reformulated, condition b) is equivalent to the
existence of orthogonal splittings $\snk$ (i.e.
$\pnk \circ \snk = id$ ) .

\be
0 \lra \widetilde V_{n,k-1} \lra \tvnk
\begin {array} {c}  \stackrel { \pnk} {\lra} \\
\stackrel{{\displaystyle \longleftarrow}}
{{\scriptstyle  \snk}} \end {array} \Vhnk
\lra 0
\ee

which are orthogonal in the sense that
\be
\snk (\Vhnk) \perp \widetilde V_{n,k-1}.
\label {sorth}
\ee
Together, these splittings give rise to an
orthogonal decomposition:
Let $\Vnk := \snk (\Vhnk)$ then
\be
\V = \perp_{n,k} \Vnk.
\label {decom}
\ee
corresponding to the double grading (\ref {dgr})
of the pathspace. The existence of such a basis,
however, is nontrivial; it is rather the crucial
point. If we look at example 1 (\ref {ex}) for
instance, we see that it satisfies conditions a)
while it fails to satisfy condition b). \\
In other words, the configuration space nature of
the pathspace leads to the prediction of a specific
orthogonal basis of $V(c,h)$.
\item [4)] An isomorphism  $\Phi '$ can be
reconstructed from the splittings $\snk$ by:
\be
\Phi' (\lii)  = \snk \circ \pnk
(\prod_{i= \inf}^1 L_i^{l_i}) , \hspace {1cm}
\mbox {for } \lii \in \Pnk
\ee
Although this isomorphism is not necessarily
admissible, it satisfies the following two
conditions
\begin {itemize}
\item [a)] $\Phi '(\Pnk) \subset \tvnk$ and
\item [b')] $\Phi ' (\Pnk)
\perp \Phi ' ({\cal P}_{n',k'})$ w.r.t.
$\lan \;  , \;  \ran_S $, if
$n \neq n'$ or $k \neq k'$.
\end {itemize}
An isomorphism satisfying the conditions a) and b')
 will be called {\bf weakly admissible}. Given a
weakly admissible isomorphism $\Phi'$, we can
always construct an admissible isomorphism $\Phi$
by orthogonalizing inside the spaces
$\Phi'(\Pnk)$. This is always possible,
since the Shapovalov form is non-degenerate and
by condition b') we already have an orthogonal
decomposition $\V = \perp_{n,k} \Phi'(\Pnk)$,
so that the form is non-degenerate on the
different parts of this decomposition; hence
diagonalisable. Independent of the chosen
orthogonalisation procedure we have:
\be
\Phi ' (\Pnk) = \Phi (\Pnk) := \Vnk.
\ee
So the spaces $\Vnk$ as well as the decomposition
corresponding to the double grading of the
pathspace $\V = \perp_{n,k}  \Vnk$ depend only on
the weakly admissible structure.
\end {itemize}

{\bf Additional remark: } In some cases (if h= 0),
it is useful to restrict oneself only to the
quasi-primary objects.
\begin {itemize}
\item [ ] Let $V_n^{q.p.} := ker (L_{-1 }
\vert_{\V_n})$,
\item [ ]      $\tvnk^{q.p.} := \ ker (L_{-1}
\vert_\tvnk)$,
\item [ ] resp. ${\widehat V}_{n,k}^{q.p.}
:= \ ker (L_{-1} \vert_{\Vhnk})$.
\end {itemize}
Analogously  we define orthogonal splittings
$s_{n,k}^{q.p.}$ by :
\be
0 \lra \widetilde V_{n,k-1}^{q.p.} \lra
\tvnk^{q.p.} \begin {array} {c}
\stackrel { \pnk} {\lra} \\
\stackrel{{\displaystyle \longleftarrow}}
{{\scriptstyle s_{n,k}^{q.p.}}} \end {array}
{\widehat V}_{n,k}^{q.p.} \lra 0 .
\ee
The whole information can be retrieved due to the
fact that the SL$(2,\C )$ sub-Verma modules based
on the $V_n^{q.p.}$ provide a basis for $\V$
\be
\bigoplus_n U(L_1) V_n^{q.p.} = \V.
\ee
The contravariance of the Shapovalov form and the
fact that $L_1 \vert_{\Pnk} \in \Pnk$ if $h = 0$
(see section 4.1) guarantee that the orthogonal
decomposition of the quasi-primary part
\be
\V^{q.p.} = \perp_{n,k}  \Vhnk
\ee
$(\Vhnk := s_{n,k}^{q.p.} (V_{n,k}^{q.p.}))$
will induce the decomposition (\ref {decom})
on the whole of $\V$.\\
This fact will be used for a construction
procedure in the next section.

\section {Admissible respresentations and
signature characters}
\subsection {Constructions for admissible
representations}

\hp One way to find an orthogonal basis of $\V$
(resp. the splittings $\snk$) with the required
restrictions is the following:

{\bf Construction 1: } We simply orthogonalize
the basis vectors of FNO with respect to the
FNO order (\ref {ord}). \\
As mentioned before, we do not know a priori
that this algorithm will work, since the Shapovalov
form is not definite and isotropic vectors may
occur. But on the other hand, the existence
(resp.\ the success of the algorithm) will be
exactly the new information gained. The
correponding isomorphism would be:
\be
\Phi_{FNO} (\lii) :=
(\prod_{i=\inf}^1 L_i^{l_i})^{\perp_{FNO}}
\label {cons}
\ee
where the superscript $\perp_{FNO}$ refers to the
above mentioned orthogonalization.
In doing the calculations, we proceeded as follows:
We just generated the Shapovalov form in the FNO
basis and used it to do the orthogonalization.
This was carried out up to $V_{32,5}$ (that is
for a 93 dimensional subspace) for the (2,5) model
and up to $V_{15,5}$ (that is for a 37 dimensional
subspace) for the (2,q odd) models for q $<35$.

We could also use any other order in which:
\be
L_{m_1} \cdots L_{m_r} \succ L_{n_1} \cdots L_{n_s}
\label {ord2}
\ee
if $r>s$. The FNO order is, however, best suited
(see section 4.2).

{\bf Construction 2: } As mentioned before, if $h$
 = 0, we can restrict ourselves to the
quasi-primary objects. Hence, we first enlarge a
basis of ${\widetilde V}_{n,k-1}^{q.p}$ which we
already know by induction to a basis of
$\tvnk^{q.p.}$ and orthogonalize the new vectors
in an arbitrary fashion. \\
This procedure can be refined in the sense that
the FNO ordering provides an even finer filtration
as (\ref {fil}) (as above, we could choose any
other ordering with the condition (\ref{ord2})),
so we can order the new basis and orthogonalize
w.r.t. this ordering as in construction 1.\\
Although this construction seems more tedious,
especially if one wants to recover the full basis,
it can be quite useful for the calculation of the
signature, since all $L_1$ descendent vectors of
quasi-primary ones have the same sign of the metric.
The calculation itself was performed in the
following manner: If h=0 we have the nice feature
that each singular vector can be taken to be of
the form:
\be
L_m L_m L_{n_r} \cdots L_{n_1} +
\mbox{ lower order terms w.r.t. the ordering
(\ref {ord})}.
\ee
If k = 2 this is especially nice since
$ V_{n,2}^{q.p.}$ is onedimensional.
The construction was carried out for $k$ = 2 up to
$n$ = 200 for all (2,q) models.

We have used both these constructions explicitly
up to different grades for n,k. The results of the
algorithms and the terms of the resulting
signature characters are contained in the next
section.

\subsection {Signature characters for the (2,q)
models}
\hp Since any admissible representation results in
an orthogonal decomposition and, in addition,
provides an orthogonal basis of $\V$ (corresponding
to the path basis), it can be used to calculate
signatures of the different parts of the
decomposition (\ref {decom}). As mentioned above,
this has been carried out for various values of n
and k. \\
For all calculated examples we find that:
\be
sign(\lan \mbox {  ,  } \ran_S )\vert_\Vnk =
(-1)^k dim \Vnk.
\ee
In other words, the Shapovalov form is $(-1)^k$
definite on $\Vnk$. This is truly a remarkable
result: from the pathspace nature we do not only
get an orthogonal basis, but we also find, and
this is very important, a basis which provides the
splitting of $\V$ into positive and negative
definite subspaces:
\be
\V = \V^+ \oplus \V^-.
\label {split}
\ee
For a discussion of the alternating sign of the
definiteness see remark in section \ref {-1^k}. \\
The data presented are far out of the region of
mere coincidence and thus lead us to the following
conjecture:

{\bf Main conjecture: } In the (2,q)-minimal
models admissible representations exist, and they
provide a splitting (\ref {split}) of $\V$ into
positive and negative definite subspaces by
\be
\V^+ := \bigoplus_{\stackrel {n}{k \; even}} \Vnk,
\hspace {2cm} \V^- :=
\bigoplus_{\stackrel {n}{k \; odd}} \Vnk.
\ee
The resulting signature character would be:

\be
\s (c(2,2N+3),h_j) (q) =  \sum_{n_1, \ldots, n_N
\geq 0} (-1)^{(\sum_{i=1}^N i n_i)} \hspace{0.5cm}
\frac {q^{N_1^2+ \cdots + N_N^2+N_j +\cdots+N_N}}
{(q)_{n_1}\cdots(q)_{n_N}}
\label {sig}
\ee

with $N_i = \sum_{j=i}^N n_j$ and $(q)_i =
\prod_{j=1}^i (1-q^j)$.

{\bf Remarks:}
\begin {itemize}
\item [1)] The signature character formulas
(\ref {sig}) have been previously conjectured by
Nahm \cite {Na}. They have been compared to the
ones given by Kent \cite {Kent1,Kent2} up to
$O(q^{100})$ for the $(2,5)$ and the $(2,7)$ model
and the two coincide \cite {Na2}.
\item [2)] The exact structure of the splittings
or of the orthogonalization is of no importance for the
induced metric on $\P$, since it is positive resp.
negative definite on each of the $\Pnk$.
So any other choice of $\Phi$ would yield the same results.
Only the weak admissibility is of importance.
\item [3)]  The characters corresponding to (\ref
{sig}) are the same except for the term
$(-1)^{\sum_{i=1}^N i n_i}$ and have been given
in \cite {FNO}. They correspond to a series of sum
rules for characters which can be interpreted in
terms of quasiparticles \cite {KRV}. In this
setting, the exponent $\sum_{i=1}^N i n_i$ is
just the sum over the number of quasi particles of
type $i$ weighted with their energy weights (see
also section \ref {particle}).
\end {itemize}

\section {Virasoro on the lattice}
\subsection {General remarks}
\label {particle}
\hp Any isomorphism $\Phi$ (\ref {iso}) is, of
course,
nothing but a realization of the Virasoro algebra
on the lattice in the sense of section 1. If we
now look at admissible representations only,
we can describe the action on the
paths themselves more explicitly. The total
structure is, however, dependent on
the specific choice of basis induced by the choice
for $\Phi$.
In general, when considering only the weakly
admissible structure of an admissible
representation, we can find the following:
\ba
L_m (\tvnk) &\subset& \widetilde V_{n+m,k+1}
\hspace {1cm} \mbox {and}
\label {inclu1}\\
L_{-m} (\tvnk)& \subset & \widetilde V_{n-m,k}.
\label {inclu2}
\ea
Furthermore, we see from the restrictions for the
annihilating ideals \cite {FNO}:
\ba
L_1^l (\tvnk ) &\subset & {\widetilde V}_{n+l,k+j}, \label {init}\\
L_2^r(\tvnk) & \subset & {\widetilde V}_{n+2r,k+N+j},
\label {diff}
\ea
for c $=$ c$(2,2N+3) $ and h = $h_j$.
\\
The first inclusion is given by the initial
condition and the second one by the difference
two condition.\\
The inclusions (\ref {inclu1}) and (\ref {inclu2})
are most easily seen in the FNO
basis:
\ba
L_m (L_{n_k}  \cdots L_{n_1}) &= & \sum_{i=k}^r
[(n_i-m) \; L_{n_k}  \cdots L_{n_i+m}
\cdots L_{n_1}]\\
&& +  L_{n_k} \cdots L_{n_{r+1}} L_m L_{n_r}
\cdots L_{n_1} ,  \mbox {where }
n_{r+1} > m \geq n_r. \nn
\ea
If any of the above vectors violates the
conditions for the FNO basis, its expression in
terms of the basis vectors contains not more
$L_i$'s than before. This is seen directly from
the annihilating ideals and the ordering as in the
proof in \cite {FNO} .\\
If we now pull the Virasoro action back with an
admissible $\Phi$ onto the pathspace, the above
inclusions
translate directly into restrictions on the $\Pnk$
\ba
L_m (\Pnk) &\subset & \P_{n+m,k} \oplus \P_{n+m,k+1}
\label {inclulm}\\
L_{-m}(\Pnk) & \subset & \P_{n-m,k-1}\oplus
\P_{n-m,k} .
\label {inclul-m}
\ea
This can be seen  in the pulled back Shapovalov
form:
First of all, we have the inclusions (\ref {inclu1})
and (\ref {inclu2}) for the
path spaces as well
\ba
L_m (\Pnk) &\subset& \bigoplus_{j=1}^{k+1}\P_{n+m,j}
 \hspace {1cm} \mbox {and}  \\
L_{-m} (\Pnk)& \subset & \bigoplus_{j=1}^{k}
\P_{n-m,j} .
\ea
Furthermore we see from the orthogonality
(\ref {decom}):
\ba
\mbox{ for  } \hii \in \Pnk & \mbox {and} &
\lii \in  \bigoplus_{j=1}^{k-1} \P_{n+m,j} \nn \\
{} \Phi^*\lan \lii,L_m (\hii)\ran_S &=&
\lan \Phi(\lii),L_m \Phi(\hii)\ran_S \nn \\
 & = & \lan L_{-m} \Phi(\lii),
\Phi(\hii) \ran_S  \nn \\
 & = & 0 \\
\mbox {by  (\ref {sorth}), since  }
L_{-m}\Phi(\lii) \in \widetilde V_{n,k-1} &
\mbox{ and} & \Phi(\hii) \in \snk (\Vhnk) . \nn
\ea
So (\ref {inclulm}) follows. (\ref {inclul-m})
follows in the same manner.
{}From the initial condition (\ref {init}) and from
the difference two condition (\ref {diff}) we find
by similar arguments:
\ba
L_1^l (\Pnk) & \subset & \P_{n+l,k} \oplus \ldots
\oplus \P_{n+l,k+j} ,\hspace {1cm} \mbox {and} \\
L_2^r(\Pnk) & \subset & \P_{n+2r,k} \oplus \ldots
\oplus \P_{n+2r,k+N+j} ,
\ea
if $c = c (2,2N+3) $ and $h=h_j$.

These restrictions concerning the action on the
paths lead us to the following additive splitting
of the usual Virasoro generators:

\ba
\label {spli}
\mbox {Let   } L_m &= &\Lhm + \Cm,
\mbox{   for } m \in {\Z}    \\
\mbox{where   }  \Lhm \vert_{\Pnk}  &=&
P_{n,k} \circ L_m \nn \\
\Cm \vert_{\Pnk}  &=& P_{n,k+1} \circ L_m.
\mbox {   for $ m >0$ }\nn\\
\Cm \vert_{\Pnk}  &=& P_{n,k-1} \circ L_m,
\mbox {   for $ m <0$ }\nn,
\ea
where the $P_{n,k}$ denote the orthogonal projection
onto $\Pnk$.
The relations (\ref {init}) and (\ref {diff}) can
now be written as:
\ba
C_1^{j+1} & =& 0 \\
C_2^{N+j+1}& =& 0, \mbox { if $c = c(2,2N+3)$ and
$h =h_j$} .
\ea
This means that the information about $h$ resp.\ $c$ is
encoded in the operators $C_1$ resp. $C_2$. These
characteristic quantities can now be simply read
off from the nilpotency index of the respective
operator.\\
We can also find commutation relations for these
operators by substituting (\ref {spli}) into the
basic Virasoro relation (\ref {comm}) and then
comparing the degrees of the various operators
w.r.t. the $K$ grading. \\
To simplify things, consider from now on $\Lhn$
and $\Cn$ for $n \in {\N} $ only and
$\Lhn^{\dag} = \widehat L_{-n} $ resp.
$\Cn^{\dag} = C_{-n}$, where the dagger
is the adjoint w.r.t. the Shapovalov form.
\begin {itemize}
\item [ ] We then have (as well as the resp.
daggered equations):
\ba
{} [ \Lhn,\Lhm ] &=& (m-n) \widehat L_{n+m} \\
{} [ \Lhn,\Cm ] + [ \Cn,\Lhm ] &=& (m-n) C_{n+m} \\
{} [ \Cn,\Cm ]&=& 0
\ea
\item [ ] and as mixed commutators if  $m \geq n$:
\ba
{} [  \Lhn^{\dag},\Lhm] + [ \Cn^{\dag},\Cm] &=&
(m+n) \widehat L_{m-n} +\d_{m,n}
\frac {c}{12}(m^3-m) \\
{} [ \Lhn^{\dag},  \Cm ] &=& (m+n) C_{m-n}\\
{} [\Cn^{\dag} , \Lhm] &=& 0
\ea
\item [] if  $m \leq n$ we obtain the daggered
equations.
\ba
{}[  \Lhn^{\dag},\Lhm] + [ \Cn^{\dag},\Cm] &=&
(m+n) \widehat L_{n-m}^{\dag} +\d_{m,n}
\frac {c}{12}(m^3-m) \\
{} [\Cn^{\dag} , \Lhm] &=& (m+n) C_{n-m}^{\dag}\\
{} [ \Lhn^{\dag},  \Cm ] &=&0.
\ea
\end {itemize}
Although these relations seem a bit more
complicated than the original Virasoro relations,
there are several reasons for the introduction of
the operators $\Lhn$ and $\Cn$. One is that in a
particle interpretation of the path space they can
be viewed as shift resp. creation-annihilation
operators.
There are several ways in which to give a particle
interpretation for the sequences $\lii$. In \cite
{KRV} for instance, a quasi particle interpretation was developed.
According to this any (2,q) model contains N different types
of quasi particles.\\
Here we shall adopt a simpler point of view. A
nonzero $l_i$ is just taken to signify the presence
of $l_i$ particles on the site $i$. The connection
to the quasi particle picture is just looking at a
quasi particle of type i as being made up by i
single particles. In this context, the spaces $\Pnk$
 are the configuration spaces
of k particles of total energy n. In these terms
the $C_n$ are one particle creation operators and
the $\Cn^{\dag}$ one particle annihilation operators
, while the $\Lhn, \Lhn^{\dag}$ conserve the total
number of
particles and thus just produce a net number of
$n$ shifts to the right resp. to the left.
In particular, we have
\ba
{} [\widehat L_1, \Lhn] &=& (n-1) \widehat L_{n+1}
\hspace {1cm} \mbox{for $i \in {\N}$} \\
{} [\widehat L_1, \Cn] + [C_1,\Lhn] &=& (n-1)
C_{n+1}.
\ea
This expresses that higher order shifts are
obtained by nested $\widehat L_1$ commutators of
$\widehat L_2$.
Furthermore, the commutators simplify  for h = 0.
Then we have from the initial condition that:
\be
C_1 = C_{-1} = 0, \hspace {1cm} \mbox{thus}
\hspace {1cm} \widehat L_1 = L_1
\ee
and the above relations simplify to
\ba
{} [ L_{\pm1}, \Lhn ] &=& (n \mp 1)
\widehat L_{n \pm 1} \\
{} [ L_{\pm1}, \Cn ] &=& (n \mp 1) C_{n \pm 1}.
\ea
Now higher order shifts and creation (resp.
annihilation)
of particles by Virasoro action are obtained by
nested $L_1$ commutators of
$\widehat L_2$ and $C_2$ (resp. $C_{-2}$).
The above remarks characterize the net action of
the Virasoro on a k-particle state.
The finer structure -how the shifts look on the
single particles or where along the chain a new
particle is created- depends on the specific
choice of basis for the isomorphism $\Phi$.
In the next section we will discuss this question
for the constructions 1 ((\ref {cons}) i.e. the
orthogonalized FNO basis) and 2 (quasi-primary
construction).

\subsection {Shifts and Creation}

\hp The simplest action of $L_n$ one could imagine
would be that $\Lhn$ shifts each particle $n$ sites
and
$\Cn$ creates a particle at site n. This is almost
the case in example 1. An action like this would be
something like the Sugawara construction
\cite {Kac}.
Problems arise, however, if the new configuration
is not allowed. Here the action becomes very
complicated. It can happen, for instance, that all
particles are shifted or more than one particle is
annihilated etc.
This is an effect induced by the nonadmissibility
of this example.
The admissibility guarantees an overall control
over the action as discussed in the previous
section.
The price paid is the "loss" of a simple shift and
creation structure.
In construction 1, however, we have further
control over the Virasoro action.
To this end we associate to a sequence $\lii$ its
finite number of nonzero members:
\be
\lii \lra (l_1^{i_1}, \ldots , l_m^{i_m})
\ee
that is the configuration with $l_k$ particles at
site $i_k$. \\
We now examine the specific action of the Virasoro
for construction 1:
\begin {itemize}
\item [1)] $\Lhn$ results in a sum of
configurations where the highest particle is
shifted at most $n$ steps. In the ordering
(\ref {ord}) we have:
\ba \Lhn (n_1^{i_1}, \ldots , n_m^{i_m}) &=&
\sum_{\mbox {config.} \in \P_{.,k}}
\l_{\mbox {config.}} \mbox {config.} \\
{} \mbox{with } \sum_i n_i = k &\mbox{and}&
\mbox {config.} \succeq (n_1^{i_1},
\ldots , (n_m-1)^{i_m},1^{i_{m}+n}).\nn
\ea
\item [2)] $\Cn$ creates a sum of configurations,
in which the new particle is at most at site n:
\ba
\Cn (n_1^{i_1}, \ldots , n_m^{i_m}) &=&
\sum_{\mbox {config.} \in \P_{.,k+1}}
\l_{\mbox {config.}} \mbox {config.} \\
\mbox{with } \sum_i n_i = k &\mbox{and}& \nn
\ea
\[ \mbox {config.} \preceq  \left\{
\begin{array}{ll} (n_1^{i_1}, \ldots ,
n_r^{i_r},1^n, n_{r+1}^{i_{r+1}},\ldots, n_m^{i_m})
 & \mbox{if  } r<n < r+1 \\
 (n_1^{i_1}, \ldots , (n_r+1)^{i_r},\ldots,
n_m^{i_m}) &\mbox{if  } r=n.  \end {array}
\right.  \]
\end {itemize}
If any of the configurations above is not allowed,
the respective coefficient is zero.
These relations are proven analogously to the
inclusions (\ref {inclu1}) and (\ref {inclu2}).
Of course, there are analogous restrictions for
the daggered operators.

{\bf Example: }
In particular in the vacuum sector of the (2,5)
model $C_2$ creates only particles at site 2.
So, for instance,
\ba
C_2 (1^2,n_2^{i_2}, \ldots , n_m^{i_m}) &=& 0 \nn \\
C_2 (1^3,n_2^{i_2}, \ldots , n_m^{i_m}) &=& 0 \nn \\
C_2 (1^4,n_2^{i_2}, \ldots , n_m^{i_m}) &=&
\sum_{\mbox {config.} \in \P_{.,k+1}}
\l_{\mbox {config.}} \mbox {config.} \nn\\
\mbox {with }&& \mbox{config.}
\preceq  (1^2,1^4,\ldots , n_m^{i_m}) \nn\\
\mbox{resp.} &&\nn\\
C_2^{\dag}(1^2,n_2^{i_2}, \ldots , n_m^{i_m}) &=&
\sum_{\mbox {config.} \in \P_{.,k-1}}\l_{
\mbox {config.}} \mbox {config.} \nn\\
\mbox {with }&& \mbox {config.}\succeq (n_2^{i_2},
\ldots , n_m^{i_m}) \nn \\
C_2^{\dag}(1^3,n_2^{i_2}, \ldots , n_m^{i_m}) &=& 0
 \nn \\
C_2^{\dag} (1^4,n_2^{i_2}, \ldots , n_m^{i_m}) &=&0.
 \nn
\ea
The relevant coefficients of this example have
been calculated up to $n$=16 and comply with the
above restrictions.

If we now turn to construction 2, we see that the
action of $L_1$ is particularly simple. In fact
one defines the whole isomorphism by the following
action:
\be
L_1  (n_1^{i_1}, \ldots , n_m^{i_m}) =
(n_1^{i_1}, \ldots , (n_m-1)^{i_m}, 1^{i_m+1}).
\ee
The rest of the Virasoro action is, however, very
complicated, since there are no further
restrictions apart from (\ref{inclulm}) and
(\ref{inclul-m}), so that in the generic case all
coefficients $\l_{\mbox {config.}} \neq 0$. In fact,
 we have not found any other admissible
representations with harder restrictions on the
overall action of $Vir$ than construction 1. For
constructions resulting from a different ordering
this is fairly easily seen.

\subsection {Nonunitarity and the pathspace metric}
\label {-1^k}
\hp If we regard the adjoint $\ddag$ with respect to
$(\; , \; )_{\P}$ we have no nice formulas for
$L_n$. This is a general fact which results
from the negative c values. It has been remarked
several times \cite {riggs,T,FT,Kac} that one can
define a positive definite scalar product on the
configuration space, but that in this product the
Virasoro algebra is not self-adjoint. In fact, it
cannot be due to the negative c. The complex
structure of the shifts can also be explained in
this way. For the treated c values one cannot find
a Sugawara construction in which the Heisenberg
algebra and the Virasoro algebra are both
self-adjoint \cite {Kac}. From the main conjecture,
we have however:
\ba
\Lhn^{\ddag} &=& \widehat L_{-n}
\label {definite} \\
C_n^{\ddag} &=& - C_{-n}.
\ea
These equations establish the connection between
the positive definite scalar product on the
configuration space and the way the Virasoro algebra
behaves under adjungation.
 In this way, the above relations can also be taken
to be a reformulation of the main conjecture. In the
particle language, this reads in the following way:
The nonunitarity of the (2,q) models manifests
itself only in the different sign between the
creation and annihilation operators under
adjungation.

\Rem In a way this is the simplest possible
nonunitarity. If we want to have the pulled
back Shapovalov form to be definite (positive or
negative) on the configuration space of k particles
$\oplus_n \Pnk$ then the definiteness changes sign
from the space of k particles to the space of k+1
particles. Let $\lii \in \Pnk$ and $ m \gg n$. From
our assumption of definiteness we have (\ref
{definite}) and $C_n^{\ddag} = \pm C_{-n}$ so:
\ba
(\lii, [L_{-m},L_m]\lii)_{\P} &=&
2mn+\frac {c}{12} (m^3-m) < 0 \\
 &=& (\lii, L_{-m}L_m\lii) \\
 &=& (\lii, \widehat L_{-m} \Lhm\lii)_{\P}
+ (\lii,C_{-m} \Cm\lii)_{\P }\\
 &=& (\Lhm \lii,\Lhm \lii)_{\P}
\pm (\Cm \lii,\Cm \lii)_{\P},
\ea
so the second term must be negative, so we must
have $C_n^{\ddag} = - C_{-n}$.

\section {Conclusion and outlook}
\hp We investigated the action of the Virasoro
algebra
on the pathspace given by the ${\cal A}_{N+1}$
graphs.
These spaces are related by a pathspace
isomorphism to the configuration space of the FB
models on coxeter $A_{q-1}$ graphs whose critical
indices are those of the (2,q) models $(q=2N+3)$.
A detailed study of the spaces under consideration
leads us to the central notion of an admissible
representation. Explicit constructions for such
representations were given. The finite dimensional
results concerning these representations lead us
to conjecture signature character formulas for
the above stated models as well as the respective
decomposition of $\V$ into positive and negative
definite subspaces. Furthermore, the explicit
Virasoro action corresponding to an admissible
representation was studied. As a result, we find
that the action of  $L_n$ can be described by a
shift
($\Lhn$) and a creation $(\Cn)$ operator. The
parameters N and j which specify the central
charge $c$ and the lowest weight $h$ of the theory
appear as nilpotency indices for these operators.
Finally, it was shown that the degree of
nonunitarity resp. of the non-selfadjointness of
the Virasoro algebra of the (2,q) models can be
understood in terms of the formulas for the adjoint
operator of $\Lhn$ and $\Cn$.\\
These results establish a close connection between
integrable models and the CFTs considered here.
On the level of the graphs the pathspace
isomorphism between the fusion graph ${\cal A}$
explains the appearance of the (2,q)-series in the
$A_{q-1}$ models. From the interplay between the
two perceptions of the objects as belonging to CFT
or statistical mechanics we gain insight into their
structure. From the CFT side we learn how the
Virasoro algebra operates as a sum of shifts and
creation. The integrable model side provides us
with signature characters and their corresponding
decompositions into positive and negative definite
subspaces.

We hope that in this spirit we can learn more
about the connection between CFT and integrable
models. There are also other series of CFT's in
which graphs appear in sum rules for characters
\cite {KRV}. A similar treatment would be of
interest. Perhaps there is also a connection to
the quasi particle interpretation of sum rules
\cite {KeMc,KeKla1,KeKla2}. In this way, one can
maybe also understand the implications of the
simple structure of the signature characters and
their role in statistical mechanics.

\section {Acknowledgements}

\hp I would like to thank W.Nahm for suggesting
this problem
to me. I also thank M. R\"osgen for the careful
reading of the manuscript. Finally I thank the
 "Studienstiftung des deutschen Volkes" for
financial support.


\begin{thebibliography} {99}

\bibitem {PP} A.Z. Patashinskii and V. L.
Pokrovskii, Fluctuation Theory of Phase Transitions,
 Pergamon Oxford 1979
\bibitem {BPZ} A.A. Belavin, A.M. Polyakov and A.B.
 Zamolodchikov, Nucl. Phys. B 241 (1984)
{\it 333}
\bibitem {P1} A.M. Polyakov, ZhETF Lett. {\bf 12}
(1970) {\it538}
\bibitem {Mig} A.A. Migdal, Phys. Lett. {\bf 44}B
(1972) {\it112}
\bibitem {P2} A.M. Polyakov, ZhETF Lett. {\bf 66}
(1974) {\it 23}
\bibitem {ABF} G.E. Andrews, R.J. Baxter and P.J.
Forrester, J. Stat. Phys. {\bf 35} (1984) {\it 193}
\bibitem {Huse} D. Huse, Phys. Rev. B {\bf 30}
(1984) {\it 3908}
\bibitem {FQS} D. Friedan, Z. Qiu and S.H. Shenker,
Phys. Rev. Lett. {\bf 52} (1984) {\it 175}
\bibitem {riggs} H. Riggs, Nucl. Phys.  B
{\bf 326 } (1989)  {\it 673}
\bibitem {pcomm} V. Pasquier, Comm. Math. Phys.
{\bf 118} ( 1988) {\it 35}
\bibitem {tlj1} H. Saleur, in Knots, Topology
and QFT, Proceedings of the Johns Hopkins Workshop,
Florence 1989, L. Lusanna ed.,
World Scientific, Singapore 1989
\bibitem {tlj2}  A. Connes, D.E. Evans,
Comm. Math. Phys. {\bf 121} (1989) {\it 507}
\bibitem {OB} A.L. Owcarek and R.J. Baxter,
J. Stat. Phys.  {\bf 49} (1987) {\it 1093}
\bibitem {Detal} E. Date et al., Phys. Rev. B
{\bf 35} (1987) {\it 2105}
\bibitem {pjourn} V. Pasquier, J.Phys. A
{\bf 20} (1987) {\it 5707}
\bibitem {PS} V. Pasquier and H. Saleur,
Nucl. Phys.  B {\bf 330} (1990) {\it 532}
\bibitem {DFJMN} B. Davies et al., Comm. Math.
Phys. {\bf 151} (1993) {\it 89}
\bibitem {JMO} M. Jimbo, T. Miwa and M. Okado,
 Lett. Math. Phys. {\bf 14} (1987) {\it 123}
\bibitem {DJK} E. Date et al., Nucl. Phys. B
{\bf 290} (1987) {\it 231}
\bibitem {FB} P.J.  Forrester and R.J. Baxter,
J. Stat. Phys.  {\bf 38} (1985) {\it 435}
\bibitem {pnuc}  V. Pasquier, Nucl. Phys. B
{\bf 285} (1987) {\it 162}
\bibitem {class1} A. Capelli, C. Itzykson and
J.-B. Zuber, Nucl. Phys. B {\bf280} (1987)
{\it 455}
\bibitem {class2} A. Capelli, C. Itzykson and
J.-B. Zuber, Comm. Math. Phys. {\bf 113} (1987)
{\it 1}
\bibitem {class3} A. Kato, Mod. Phys. Lett. A
{\bf 2} (1987) {\it 585}
\bibitem {class4} D. Gepner and Z. Qui,
Nucl. Phys. B {\bf 285} (1987) {\it 423}
\bibitem {KS} W.M. Koo and H. Saleur, preprint
USC-93-025/YCTP-P22-93
\bibitem {IT} H. Itoyama and H.B. Thacker,
Phys. Rev. Lett. {\bf 58} (1987) {\it 1395}
\bibitem {T} H.B. Thacker,
Physica  {\bf 18}D (1986) {\it 348-354}
\bibitem {FT} H. Frahm and H.B. Thacker,
J. Phys. A {\bf 24} (1991) {\it 5587}
\bibitem {a.r} J. Kellendonk and A. Recknagel,
Phys. Lett. B {\bf 298} (1993) {\it 329}
\bibitem {KRV} J. Kellondonk, M. R\"osgen
and R. Varnhagen,
Int. J. Mod. Phys. A {\bf 9} (1994) {\it 1009}
\bibitem {FNO} B. Feigin,  T. Nakanishi and
H. Ooguri, Int. J. Mod. Phys. A {\bf 7} suppl.
{\bf 1A} (1992) {\it 217}
\bibitem {Kac} V.G. Kac and A.K. Raina, Highest
weight rep. of infinitie dimensional Lie algebras,
World Scientific 1987
\bibitem {Na} W. Nahm, private communication
\bibitem {Kent1} A. Kent,
Phys. Lett. B {\bf 269} (1991) {\it 314}
\bibitem {Kent2} A. Kent,
Comm. Math. Phys. {\bf 143} (1991) {\it 1}
\bibitem {Na2} A. Kent, private communication
\bibitem {KeMc} R. Kedem and B.M. McCoy,
J. Stat. Phys. Vol. {\bf 71} (1993)
{\it 865}
\bibitem {KeKla1} R. Kedem et al.,
Phys. Lett. B {\bf 304} (1993) {\it263}
\bibitem {KeKla2} R. Kedem et al.,
Phys. Lett. B {\bf 307}  (1993) {\it 68}
\end {thebibliography}
\end {document}